\documentclass[onecolumn,authoryear]{els-mrw}

\usepackage{amsmath,amssymb,amsfonts,amsthm,makeidx,graphicx}
\usepackage{txfonts}
\usepackage{helvet}
\usepackage[colorlinks=true,citecolor=blue, urlcolor=blue, ]{hyperref}
\usepackage[normalem]{ulem}
\usepackage{xcolor}

\usepackage{soul}

\def\aj{AJ}
\def\apj{ApJ}

\def\apjl{ApJL}
\def\apjs{ApJS}

\def\aap{A\&A}
\def\araa{ARAA}
\def\mnras{MNRAS}

\def\pasa{PASA}
\def\pasp{PASP}

\def\na{Nature}
\def\nat{Nature}

\def\rmxaa{RevMexAA}

\begin{document}

\graphicspath{{./figures/}}

\chapter{Luminous Blue Variables}\label{ch:LBV}

\author[1]{Nathan~Smith}

\address[1]{\orgname{University of Arizona}, \orgdiv{Department of
    Astronomy \& Steward Observatory}, \orgaddress{933 N. Cherry Ave., Tucson, AZ 85721, USA}}

\articletag{LBV}

\maketitle

\begin{abstract}[Abstract]
  Luminous Blue Variables (LBVs) are a class of  massive blue supergiants exhibiting irregular and eruptive instability, sometimes accompanied by extreme mass loss.  While they have often been considered to be a brief but very important transitional phase in the evolution of massive single stars,  mounting evidence suggests that LBVs are actually a phenomenon associated with binary star evolution by way of mass gainers and mergers.  If true, this would leave single massive stars without a way to shed their H envelopes and become Wolf-Rayet (WR) stars, requiring a revision of our most basic paradigm of massive star evolution.  Understanding their properties and role in evolution is therefore extremely important.  The eruptive mass loss of LBVs is also invoked to account for some of the most extraordinary supernovae (SNe) with dense circumstellar material (CSM), which again contradicts the standard single-star scenario where LBVs are massive stars in transition to the WR phase.  This chapter reviews many aspects of LBVs including terminology and classification, the physical properties of the stars, their mass loss, and their variability,  as well as their observed companion stars, implications of their surrounding stellar populations, and their circumstellar nebulae.  These are put into context with proposed evolutionary scenarios and possible mechanisms that drive their eruptive variability.
\end{abstract}

\begin{glossary}[Glossary]
  \term{circumstellar material (CSM) -} gas and dust located from a few stellar radii out to a few pc from a star, generally composed of material that has been previously ejected by the star through winds or episodic mass loss \\
  \term{common envelope -} a phase in binary evolution when a star is engulfed by its companion's envelope; may eject the envelope and leave a close binary, or it may lead to a merger\\
  \term{Eddington limit-} tipping point where the radiation force exceeds the force of gravity; typically this refers to the ``classical" Eddington limit, which assumes electron scattering opacity\\
  \term{hypergiant  -} a star with luminosity class Ia+, showing spectroscopic signatures of very high luminosity and strong mass loss\\
  \term{Luminous Blue Variable (LBV) -} a class of H-rich blue supergiants with irregular brightness variations and strong mass loss\\
  \term{merger} when two stars merge into one, probably accompanied by a significant eruptive mass-loss event and luminous outburst\\
  \term{metallicity -} fraction of mass in elements heavier than hydrogen and helium\\
  \term{Wolf-Rayet -} a class of hot, luminous stars with strong and fast winds, producing strong emission lines; typically refers to H-poor stars that are the exposed He cores of evolved massive stars, although there are H-rich Wolf-Rayet stars as well \\ 
 \vspace{2 mm}
  \term{Wind mass loss:}\\
  \term{\hspace{5 mm}stellar wind  -} continuous steady mass loss from an individual star's surface\\
  \term{\hspace{5 mm}line-driven wind -} radiation-driven stellar wind where UV absorption lines are the primary source of opacity\\
  \term{\hspace{5 mm}dust-driven wind -} radiation-driven stellar wind where dust grains are the primary source of opacity\\
  \term{\hspace{5 mm}continuum-driven wind -} radiation-driven stellar wind where electron scattering is the primary source of opacity\\
  \term{\hspace{5 mm}super-Eddington wind -} a temporary strong  wind phase caused by the star's total luminosity rising above the Eddington limit\\
  \term{Episodic mass loss:}\\
  \term{\hspace{5 mm}eruption -} non-steady mass loss occurring during an outburst, stronger than can be explained by normal winds\\
  \term{\hspace{5 mm}giant eruption -} a particularly extreme case of an LBV eruption where substantial mass loss occurs, accompanied by a significant \\ \hspace{7 mm}increase in the star's total bolometric luminosity; the prototype is the 19th century eruption of $\eta$ Carinae\\
  \term{\hspace{5 mm}outburst -} observed temporary/irregular brightening of a star; non-supernova transient event\\
  \term{\hspace{7 mm}S Dor outburst -} a type of irregular brightening characteristic of LBVs, thought to be caused by a cooler expanded photosphere, \\ \hspace{7 mm}shifting the bolometric flux from the UV to visual wavelengths; named for the variations of the prototype LBV star in the Large \\ \hspace{7 mm}Magellanic Cloud \\
  \term{\hspace{5 mm}supernova impostor - } a term for extragalactic non-SN transients; often used interchangeably with LBV giant eruption\\
  \term{Supernova (SN):}\\
  \term{\hspace{5 mm}pair instability  supernova (PISN) -} a very luminous thermonuclear SN arising during late-phase burning stages of very massive stars; \\ \hspace{7 mm}leaves no compact remnant \\
  \term{\hspace{5 mm}pulsational pair instability  (PPI) -} a partial explosion that results from explosive nuclear burning in late phases, which does not \\ \hspace{7 mm}destroy the star, but results in repeating shell ejections that collide; leaves a black hole remnant \\
  \term{\hspace{5 mm}core-collapse supernova -} luminous transient from a terminal explosion of a massive star\\
  \term{\hspace{5 mm}SN~Ic -} supernova with spectrum showing neither helium nor hydrogen lines\\
  \term{\hspace{5 mm}SN~Ib  -} supernova with spectrum showing helium but not hydrogen lines\\
  \term{\hspace{5 mm}SN~IIb -} like SN~Ib, but with fleeting H lines, death of a star with a small residual H envelope\\
  \term{\hspace{5 mm}SESN   -} stripped-envelope supernova; Types Ic, Ib, and IIb together, arising from stars that have lost most or all of their H envelopes\\
  \term{\hspace{5 mm}SN~II -} supernova with spectrum showing hydrogen lines;  usually refers to Types II-P and II-L together\\
  \term{\hspace{5 mm}SN~IIn-} SN~II with narrow emission lines, thought to arise from interaction with dense CSM\\
  \term{\hspace{5 mm}SN~Ibn -} SN~Ib with narrow emission lines, thought to arise from interaction with dense H-poor CSM\\
 
\end{glossary}

\begin{glossary}[Nomenclature]
  \begin{tabular}{@{}lp{34pc}@{}}
    LBV & Luminous Blue Variable\\
    BSG & Blue supergiant\\
    RSG & Red supergiant\\
    YHG &  Yellow hypergiant\\
    WR & Wolf-Rayet star\\
    WN, WC, WO & Wolf-Rayet subtypes that are evidently N-rich, C-rich, or O-rich, respectively\\
    WNH & the class of N-rich WR stars that still have significant H envelopes; may still be on the main-sequence\\ 
    ILOT & Intermediate-luminosity Optical Transient\\
    LRN & Luminous red nova\\
    SN & Supernova\\
    SLSN & Superluminous supernova \\
    PISN & Pair  instability supernova \\
    IMF & Initial mass function; distribution of initial masses of stars\\
    LMC, SMC & The Large and Small Magellanic Clouds\\
    CSM & Circumstellar material\\
    RLOF & Roche Lobe Overflow in a binary system, causing mass transfer to a partner or mass loss from the system
\end{tabular}
\end{glossary}

\begin{BoxTypeA}[]{Key points}
\begin{itemize}
\item LBVs are the most luminous supergiants in any star-forming galaxy, and they have the highest mass-loss rates of any class of stars.
\item LBVs are simultaneously thought to be a critically important phase in the evolution of massive stars, but also very poorly understood in terms of their evolutionary origin, evolutionary phase, and the physics governing their instability and strong mass loss.
\item{Current evidence disfavors the older view that LBVs represent a transitional phase in the lives of very massive single stars whose mass loss is triggered by an unknown instability, and instead, evidence points toward LBVs and their eruptions as an observed phenomenon associated with violent binary interaction, including rapid mass transfer and stellar merger events.  In this scenario, LBVs are essentially massive blue stragglers. There is much debate about this.}
\item{The extreme giant eruptions of LBVs, in particular, are problematic for the older view of LBVs.  In the massive single star model, the eruptive mass loss is triggered by the star exceeding the Eddington limit, but there has never been any physical explanation for why this happens or what causes the dramatic increase in luminosity.  In a binary model, the energy of a giant eruption is naturally explained as the release of orbital energy in a stellar merger.}
\item{The strong winds and dense CSM of some LBVs are reminiscent of the dense CSM seen around some classes of interacting SNe, such as SNe~IIn and SNe~Ibn, suggesting that some LBVs may be a terminal rather than a transitional phase of evolution.}
\end{itemize}
\end{BoxTypeA}

\section{Introduction}\label{intro}

Luminous Blue Variables (LBVs) are the most luminous supergiants in any star-forming galaxy, and they have the highest observationally inferred mass-loss rates of any class of stars.  Because of their strong mass loss --- often in the form of sporadic eruptions that vastly exceed the mass-loss rates of normal winds --- they are thought to be critical for understanding the evolution and fates of massive stars.  This eruptive mass-loss (which may or may not remove the H envelope) profoundly influences the fate of the star and the type of eventual supernova (SN) explosion \citep{so:06,smith:14}. The way LBVs actually figure in this evolution and the physical mechanisms of their outbursts remains challenging to understand, and controversial.  The specific role of this eruptive mass loss in stellar evolution is important to understand, since unlike normal stellar winds, it is likely to be insensitive to metallicity \citep{so:06}.  Stellar evolution models that include eruptive mass loss would therefore scale quite differently to lower metallicity regimes in the early universe, as compared to models with normal metallicity-dependent winds \citep{heger:03}.  Moreover, LBVs are thought to be related to some extragalactic non-supernova (SN) transients \citep{smith:11lbv,vdm:12} and their mass-loss is reminiscent of extreme pre-SN eruptions \citep{mauerhan:13,smith:14ip,sm:07,smith:07,gal-yam:07}.  Understanding the origin of the LBV instability, the evolutionary scenario of LBVs, their larger role in massive star evolution, and their connection to extragalactic transients and some classes of SNe remains an enduring challenge.

LBVs were recognized early-on as the brightest blue irregular variables in nearby spiral galaxies like M31, M33, and NGC~2403 \citep{hs:53,ts:68}, originally referred to as the ``Hubble-Sandage variables''.  Most of these were discovered in the course of photometric monitoring studies of Cepheids, and for a time, it was hoped that LBVs and their variability might similarly be used as distance indicators, although the erratic nature of LBVs made this infeasible.  Later, it was recognized that famous Galactic objects like P~Cygni and $\eta$ Carinae had spectacular outbursts in the 17th and 19th centuries, respectively, and appeared to share many of the same properties. These blue supergiant irregular variables were therefore grouped together into a larger class, and the name ``LBV'' was introduced by \citet{conti:84}.  Some researches have expressed dismay that LBVs are poorly defined, but this may have been intentional from the outset.  Paraphrasing \citet{conti:84}, LBVs were envisioned as an ``other" category --- essentially any evolved hot supergiant that wasn't an O-type star or a  Wolf-Rayet (WR) star.\footnote{Somewhat amusingly, in an apparent contradiction, some LBVs are actually both O-type stars and WR stars.  Namely, several of the more luminous LBVs, when in their quiescent state between eruptions, exhibit a hybrid Ofpe/WN9 spectral type.} Given their diversity, it may be helpful to think of LBVs not so much as a well-defined class of variable star, but as an observational phenomenon associated with irregular variability and strong mass loss that may be related to an as-yet unidentified eruptive instability or other trigger.  There were efforts in the 1980s and 1990s to more rigidly define LBVs, inevitably leading to several sub categories populated by only a few objects each \citep{hd:94,vg:01}.  Over a dozen LBVs are now identified in the Milky Way (MW) and the Large and Small Magellanic Clouds (LMC/SMC), and a similar number resides in other Local Group galaxies \citep{hd:94,vg:01,smith:04,clark:05,st:15}.  Stars that spectroscopically resemble LBVs with similar luminosity and color, but which have not (as yet) been observed to show the eruptive variability of LBVs, are often called ``LBV candidates''; these are presumed to be temporarily dormant LBVs. In the past few decades, many LBV candidates have indeed become confirmed LBVs by displaying irregular outbursts \citep{clark:03,clark:05,valeev:09,buemi:11,miro:14,ghg:15,sholukhova:18,smith:20,ss:21,mojgan:22,martin:23}.

Perhaps the LBV grab-bag was inclusive because \citet{conti:84} recognized that despite the diversity and individuality in this hodge-podge collection of stars, with each one seeming unique at some level, they nevertheless might perform a similar role in evolution.  That evolutionary role envisioned by Conti and others was that LBVs provided the much-needed service of removing the H envelope, which was required to turn massive O-type stars into H-poor WR stars.  This sequence, where single massive stars evolve to become WR stars by virtue of their own strong mass loss, became a central paradigm of massive star evolution, and is often referred to as the ``Conti scenario" \citep{conti:75,maeder:96}.  LBVs were the lynch-pin of this scenario, since they were the only game in town in terms of observed stars with high-enough mass-loss rates over most of the relevant stellar initial mass range.  The observed lack of red supergiants (RSGs) at high luminosity \citep{hd:79}, and the tendency of contemporary 1-D stellar evolution codes to crash when evolved off the main sequence with H envelopes in tact, made this appealing.  Moreover, this scenario was straightforward to accomplish in 1-D stellar evolution codes, by simply invoking outrageously high mass-loss rates for a short time until the H envelope was gone.  In this role, LBVs are analogous to the thermal-pulsing AGB phase of evolution in lower-mass stars, except that instead of removing the H envelope to reveal a degenerate white dwarf, LBVs remove the H envelope to reveal the non-degenerate He core that is seen as a WR star.

In a previous review about LBVs, \citet{smith:17} summarized the traditional view of LBVs from the 1980s and 1990s (see, e.g.,  \citealt{hd:94}, \citealt{vg:01}, and many contributions in \citealt{nota97}) according to four hallmarks that have been used to define LBVs:  (1) Their evolutionary stage is a transitional stage in the lives of single very massive stars between O stars and WR stars, when the H envelope is lost by strong mass loss; (2) typical LBV variations (also known as S Doradus variations) of ``classical LBVs'' are changes in apparent temperature at constant bolometric luminosity, caused by an expanded pseudo-photosphere (which in turn results from a very large increase in mass-loss rate), where the star shifts from its hot quiescent state on the S Dor instability strip over to the cooler constant temperature strip at around 8,000 K; (3) this strong eruptive mass loss halts their redward evolution, explaining the missing RSGs at high luminosity and the formation of WR stars; and (4) some LBVs suffer even more dramatic `Giant Eruptions' like the 19th century eruption of $\eta$ Car, when much larger amounts of mass can be lost, with these being instrumental in forming WR stars.  

This scenario presented a fairly clean and simple view of massive star evolution and the role of LBVs in that evolution.  However, as noted previously by \citet{smith:17}, most of these hallmarks are unraveling and have not stood the test of time.  Further study in the past two decades has given rise to a number of ways that the observed properties of massive stars and SNe contradict this scenario, as described below.  This has given rise to a central debate concerning LBVs and their role in stellar evolution.  Since LBVs are thought to be the main agent responsible for removing the H envelope to make WR stars and stripped-envelope SNe (SESNe), this debate is critical for understanding massive star evolution more generally.  In the simplest terms, the debate centers on the following question:   Is the LBV phase really an important transitional phase that {\it all} very massive stars pass through on their way to the WR phase, or is it a phenomenon exclusively associated with binary evolution? Knowing the answer is very important, because if LBVs are a binary product, then they must be removed from the single-star evolutionary scenario; doing so, however, prohibits single massive stars from becoming WR stars throughout the relevant range of initial masses.  This, in turn, raises basic questions about the fates of massive single stars.

This chapter will review the definition and properties of LBVs, their various manifestations among nearby stellar populations and transients, their circumstellar material (CSM) that is evidence for past mass loss, their environments and implications for single or binary evolution, and possible explanations for the origin of their instability and strong mass loss.

%We first rrs (\Secref{sec:intro_theory}). 
%\section{Theoretical overview of pair instability}\label

\section{Terminology and Observed Properties}\label{term}

\subsection{Terminology}

First, a disclaimer: the author can't control how the term LBV is used, and isn't trying to advocate that the way it is used is good or bad.  But it is useful to acknowledge that the use of ``LBV" in the literature has been rather inconsistent.  A critical way to view this is that there is confusion about what is a ``true" LBV.  A more constructive view might be that some of the rigid definitions introduced for LBVs are too narrow for the actual observed phenomena, to the point that they hinder our understanding of the broader phenomenon.  Many of the supposedly defining traits of LBVs are based on just a few prototypes, but as more LBV-like stars are discovered or studied in more detail, those stars often fail to obey all the rigid definitions.  When confronting new discoveries that don't fit, one is faced with the choice of just saying that those objects aren't true LBVs (and if not LBVs, then what are they?), or admitting that the rigid criteria based on a few objects may be inadequate.  Researchers tend to disagree on these points, and inconsistent usage therefore proliferates in the literature.  This section will attempt to gather and discuss the various ways that the term LBV has been used.

There are three main ways that the term LBV is applied in the literature:

{\bf 1. Stars.} First, we have the classical LBVs, sometimes referred to S Doradus variables or Hubble-Sandage variables.  These are luminous hot supergiants/hypergiants in the Milky Way and a few nearby galaxies that exhibit the typical irregular variability that is the hallmark of the original LBVs \citep{conti:84,hd:94,smith:04}.  Typically, these will show cooler temperatures as they brighten, leading to the presumption that their brightening is mainly due to a redistribution of bolometric luminosity from the UV to the optical.  Classic examples are AG Car, HR Car, R127, and S~Dor.  In this category, we might also include bright blue irregular variables that don't quite obey the defined criteria to make them S Dor variables, such as the erupting stars in M33 that brighten at constant temperature (see below).  Finally, we might also include nearby stars that resemble classical LBVs spectroscopically, but have not yet demonstrated the same variability.  These are the aforementioned LBV candidates.  Readers should be aware that some authors drop the term ``candidate" when discussing these together with the rest of the class.

{\bf 2a.  Giant Eruptions.}  LBVs are also identified by their pronounced increases in luminosity and signs of eruptive mass loss.    Famous examples are the giant eruptions of $\eta$ Carinae in the 19th century and P Cygni in the 17th century, as well as HD~5980 in the SMC in the 1990s.  These giant eruptions are often assumed to require a true increase in bolometric luminosity \citep{hds:99}, unlike S Doradus events, although for some objects like P Cyg and HD~5980, it remains unclear how much the bolometric luminosity actually increased.  Due to the eruptive mass loss, many LBVs are surrounded by fossil shell nebulae that are N-enriched, again with the Homunculus of $\eta$ Carinae and its more extended debris being the best studied example.  Many LBV candidates are surrounded by such N-rich shells, thought to be the result of past eruptions, even if the bright outbursts were not recorded.

{\bf 2b. Extragalactic transients / SN impostors.} Modern surveys of the transient sky continue to find many examples of bright transients that (sometimes after additional study) do not appear to be terminal SNe.  These transient sources, with peak absolute magnitudes ranging from roughly $-$10 to $-$15 mag (brighter than novae, and fainter than SNe), were sometimes therefore called ``SN Impostors" \citep{vandyk:00,vdm:12}.  Their spectra vaguely resemble LBVs with strong H emission lines that typically have widths of a few 10$^2$ km s$^{-1}$ \citep{smith:11lbv}, much narrower than the broad lines in most SNe II.  Since LBV giant eruptions were a well-established precedent for non-SN brightening events, many of the early transients of this sort were interpreted as LBVs.  The reason these are considered here somewhat separately from LBV giant eruptions is that for SN impostors, we usually know much less (if anything)  about the progenitor star or the putative survivor.  Some SN impostors may be terminal events after all, since their very faint and dust enshrouded states are difficult to constrain in distant galaxies \citep{kochanek:12}.    Famous examples are the aptly named SN~1997bs, as well as SN~1954J, SN2000ch, and others \citep{vandyk:00,wagner:04,smith:01,smith:11lbv,kochanek:12,vdm:12}.  As many more have been discovered in modern transient searches, it has opened a Pandora's Box with transient sources that have exhibited an increasingly wide variety of peak luminosity, decline rates, and spectral evolution.  Upon discovery, many of these are initially classified as LBVs or SN impostors by transient observers, even though later consideration may lead authors to favor a different classification. Other overlapping names have been introduced for these non-SN transients, including Intermediate Luminosity Optical Transients (ILOTs), Luminous Red Novae (LRNe), stellar mergers or mergerbursts, and others.  

{\bf 3. Supernova progenitors.}  Lastly, the term LBV is often used to describe some SN progenitors, especially progenitors of Type~IIn or Ibn events.  This includes (1) rare cases of SN progenitors like SN~2009ip, SN~1961V, or SN~2006jc that are caught in one or more LBV-like pre-SN eruptions \citep{smith:10ip,mauerhan:13,goodrich:89,pastorello:07}, (2) pre-SN detections of progenitors that (while may be seen in only a single epoch and lack demonstrated variability), are nevertheless very luminous and interpreted as being consistent with known LBV stars \citep{gal-yam:07,gl:09,smith:11jl}, (3) progenitors of SNe that are inferred to have LBV-like variability in their wind based on observed diagnostics of CSM interaction in the SN \citep{kv:06,trundle:08}, and (4) SNe~IIn and super-luminous SNe~IIn that require very extreme eruptive pre-SN mass loss (i.e., losing several M$_{\odot}$ in decades or less), for which there is no observed precedent besides LBV giant eruptions \citep{sm:07,smith:07,smith:08tf,smith:10,smith:14ip,smith:14,smith:24,miller:10,fox:10,kiewe:12,kilpatrick:18,tartaglia:20,bilinski:24,dickinson:24,dukiya:24}.

Readers, especially students and others new to LBVs, should be cautiously aware that the use of the term LBV in the literature covers a broad range of objects that might be grouped together loosely as the LBV phenomenon, which arguably maintains the spirit of the original invention of the term by \citet{conti:84}.  At the same time, readers should be cognizant of the possibility that an object called an LBV in the context of pre-SN mass loss is not necessarily the same thing as a classical S~Dor variable.  In the case of pre-SN mass loss, the term ``LBV" may be interpreted as shorthand for a H-rich supergiant with really strong eruptive mass loss that far exceeds any known steady wind, and with bulk outflow speeds similar to LBVs. For example, a red or yellow supergiant that suffers some sort of explosive pre-SN mass loss that has accelerated outflow speeds and mass-loss rates much higher than a typical red or yellow supergiant wind might look, to SN observers, a lot like the outflows from LBVs, even if the progenitor star might not have been called an LBV if we could see it up close.  Since we still do not know what triggers the instability or extreme mass loss of LBVs, and since LBVs themselves exhibit considerable diversity, perhaps using LBV as an umbrella term need not be too confusing or distressing.

\begin{figure}[htbp]
  \centering
  \includegraphics[width=0.73\textwidth]{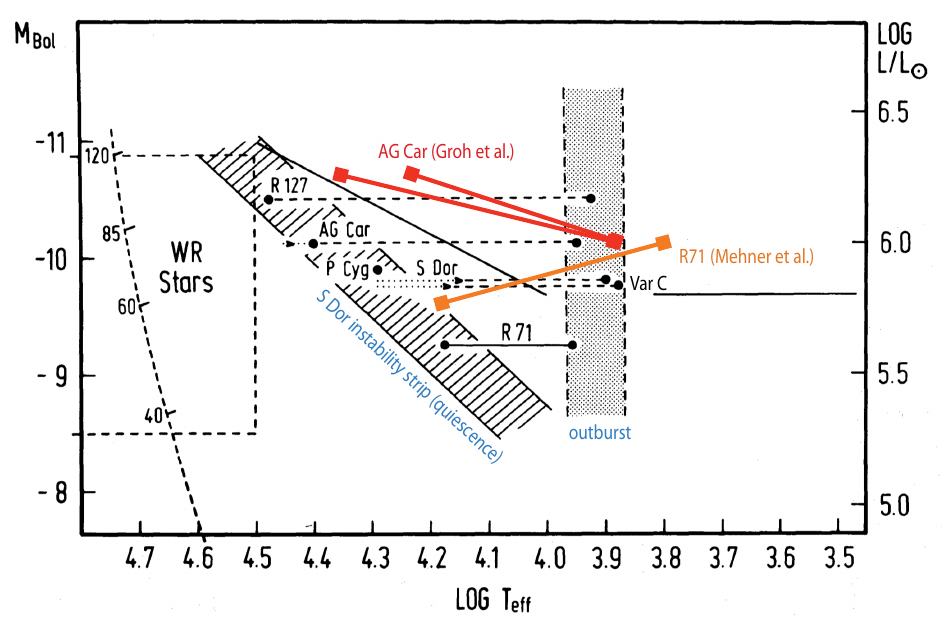}
  \caption{This is the original plot (adapted from \citealt{wolf89}, with blue, orange, and red annotations added by the author) showing the HR diagram locations proposed by \citet{wolf89} for LBVs at quiescence (the S Dor instability strip) and in outburst (constant temperature strip), including L and T estimates for R127, AG Car, P Cyg, S Dor, Var C in M33, and R71 (black).  More recent estimates (based on detailed analysis with modern CMFGEN models) for AG Car (red; \citealt{groh09}) and R71 (orange; \citealt{mehner:13,mehner17}) are also shown for comparison.  The two different quiescent states for AG Car are for two different quiescent periods when it had different temperatures \citep{groh09}.  The point here is that the location of the S Dor instability strip and the idea that S Dor outbursts occur at constant L$_{\rm Bol}$ are both somewhat suspect based on modern data.}
  \label{fig:sdor}
\end{figure}

\subsection{Observed properties}

\subsubsection{Luminosities and temperatures}

As noted above, a traditional hallmark of classical LBVs concerns their locations on the HR diagram during their outburst variability cycle.  When in their quiescent state between eruptions, LBVs are usually presumed to reside along the diagonal ``S Doradus instability strip'', first proposed by \citet{wolf89}. During standard LBV outbursts (also known as S Doradus variations) they brighten significantly, thought to arise from changes in apparent temperature occurring at roughly constant
bolometric luminosity.  As their photospheres expand and their apparent temperatures cool to around 8,000 K, their distribution of bolometric flux shifts to shorter wavelengths, causing the observed visual-wavelength brightening.  The S Dor instability strip and these constant $L_{\rm Bol}$ excursions are shown in Figure~\ref{fig:sdor}. 

The diagonal S Dor strip and the vertical outburst strip are often reproduced in the literature as the defining locations of LBVs, with their hot and cool states connected by horizontal lines.  However, these locations and the horizontal lines connecting LBVs in quiescence and outburst should be taken with a large grain of salt.   The original S~Dor strip was defined by only 6 LBVs with L and T estimated mostly from photometry --- yet two of these (S~Dor and Var~C) typically have cooler temperatures in quiescence and are usually not found on the S Dor strip (noted by their dashed left locations in Figure~\ref{fig:sdor}), and one of these (P Cygni) is not really an S Dor variable (it is considered an LBV because of its 17th century giant eruption; it has not shown the classic S Dor-like variability since then).  Moreover, physical parameters like L and T are difficult to estimate for LBVs due to their strong winds that mask the normal spectral type and affect broadband fluxes, making L and T very uncertain from optical photometry.  For comparison, Figure~\ref{fig:sdor} also includes more modern estimates of L and T derived from state-of-the-art radiative transfer models compared to high-quality spectra for two classic LBVs from this original plot.  AG~Car is shown in red \citep{groh09} and R71 in orange \citep{mehner:13,mehner17}.  With more recent analyses, neither of these seem to vary at constant $L_{\rm Bol}$ (in fact, showing opposite trends) and their quiescent locations differ significantly from the original plot, with R71 at the edge of the S~Dor instability strip, and AG~Car residing off the strip.  \citet{wolf89} may have drawn the S~Dor instability strip in a very different location, had he seen these updated results.

Moreover, values for the luminosity obviously also depend on adopted distances and reddening.\footnote{Interestingly, it was an early hope that the diagonal S~Dor instability strip and an amplitude-luminosity relation might provide a reliable distance indicator \citep{wolf89}, since LBVs are even brighter than Cepheids.  However, the outbursts of LBVs turned out to be too erratic, and multiple outbursts for the same LBV would yield different amplitudes and distances.}  While distances and local reddening for extragalactic LBVs in the LMC/SMC or M33 are fairly reliable, the distances for Milky Way LBVs have been highly uncertain.  The literature contains many plots of the HR diagram showing that LBVs cluster tightly along the S~Dor strip, but this may be quite misleading.  In trying to track down prior LBV distance estimates from the literature, \citet{smith:19} found that many early studies of individual LBVs had very uncertain distances from a variety of methods (such as the distance-reddening relation), and for the sake of comparison, many authors chose to adopt a seemingly plausible distance value that landed an LBV on the S Dor instability strip.  These distances were then often cited by subsequent authors, who then used them to show LBVs landing on the S Dor instability strip.  This sort of circular reasoning should give one pause about the reality of the S Dor instability strip.  \citet{smith:19} then adjusted literature values of the luminosity of each Galactic LBV using the new Gaia DR2 distance, finding that most Milky Way LBVs do not reside on the S Dor instability strip, with some way below or far above it.  While the error bars on Gaia DR2 distances are quite large and this should be updated\footnote{\citet{mahy:22} provided a number of Gaia DR3 distances fo LBVs; a few were significantly different, but many were consistent with the DR2 values.}, this DR2 study was the first collection of direct distances to Milky Way LBVs, and from this, the S Dor instability strip does not appear to be a real feature of Galactic LBVs.   Readers should regard the location and reality of the S Dor instability strip with some caution, especially if one wishes to connect potential theoretical explanations for the LBV instability \citep{lf:88,gk:93,sc:96,lamers:97,guzik:99,stothers:99,grafener:12,owocki:13,owocki:15,grafener:15,jiang:15,jiang:18,grass:21,agrawal:22,cheng:24} to match their locations on the HR diagram.

\begin{figure}[htbp]
  \centering
  \includegraphics[width=0.95\textwidth]{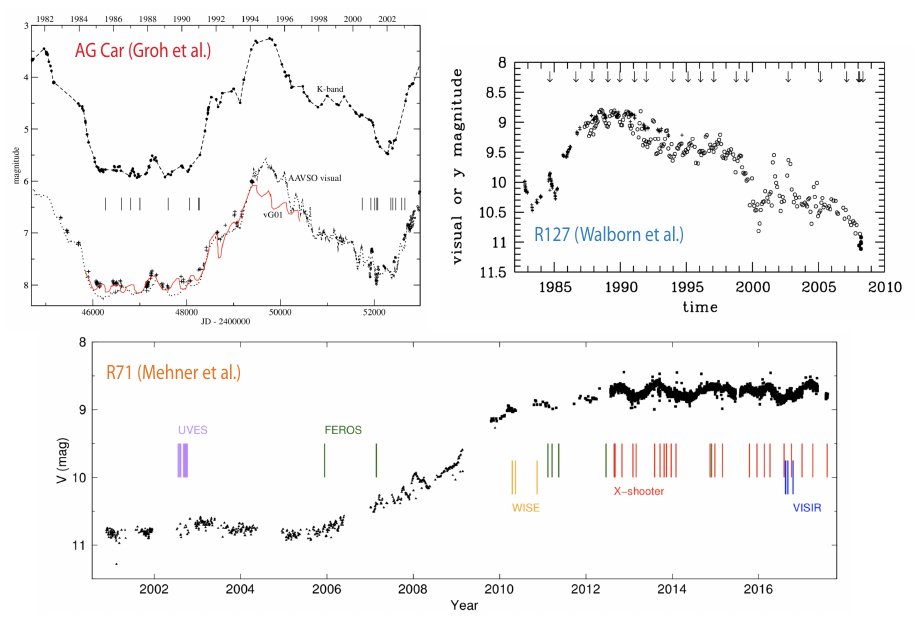}
  \caption{Examples of S Doradus variations in three classic LBVs.  Upper left:  AG Carinae \citep{groh09}.  Upper right: R127 \citep{walborn:08}.  Bottom: R71 \citep{mehner17}.}
  \label{fig:sdorlc}
\end{figure}

\subsubsection{Variability}

{\it S Dor variations:}    The most common type of variability exhibited by LBVs is seen as irregular brightening phases that occur on timescales of years to decades, although sometimes also showing considerable structure on shorter timescales during an event (see Fig.~\ref{fig:sdorlc}).  These outbursts are not periodic, and each outburst is somewhat different from the last.  As noted above, a traditional hallmark of LBVs is that their normal S~Dor brightening events occur when the star cools at roughly constant $L_{\rm Bol}$, therefore shifting much of its flux from the UV into the optical.  Since there is a temperature-dependent location for LBVs in quiescence (the S Dor instability strip) and since the peak of the outbursts exhibit similar apparent temperatures, there is expected to be a relationship between the amplitude of $B$-band or $V$-band brightening and the true luminosity, for the simple reason that the hotter and more luminous LBVs have a bigger bolometric correction.  S Dor variations typically have $B$-band amplitudes ranging from a few tenths of a magnitude for lower-luminosity LBVs, up to $\sim$2.5 mag for the more luminous LBVs \citep{wolf89,vg:01}.  The idea that these variations occur at constant $L_{\rm Bol}$, which was inferred initially from photometric colors and assumed bolometric corrections, has not held up under more detailed scrutiny with quantitative analysis.  There are several examples of LBVs that have had their fundamental parameters estimated with sophisticated radiative transfer codes like CMFGEN, and a number of studies find that an LBV's value for $L_{\rm Bol}$ can vary substantially (see Fig.~\ref{fig:sdor}) during an S~Dor phase, sometimes with L increasing or decreasing relative to the quiescent value \citep{groh09,mehner17,vg:01,clark:09}.  Admittedly the inferred changes in $L_{\rm Bol}$ are not huge (roughly a factor of 2; but again, going either up or down), but still worth considering when thinking about mechanisms for the LBV instability.

\begin{figure}[htbp]
  \centering
  \includegraphics[width=0.85\textwidth]{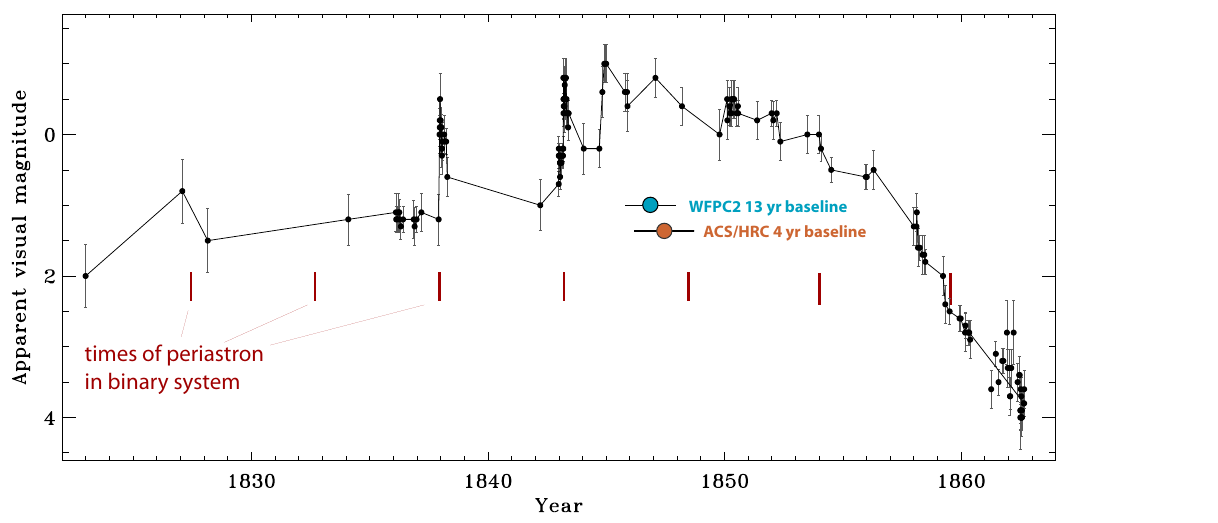}
  \caption{The visual-wavelength light curve of the 19th century Great Eruption of $\eta$ Carinae, using historical photometry from \citet{sf:11}.  This figure is adapted from \citet{smith:17pm}, showing the estimated dates for the ejection of the nebula derived from {\it HST} proper motion measurements \citep{smith:17pm}, and the estimated times of periastron passes based on the current binary system \citep{sf:11}.}
  \label{fig:etalc}
\end{figure}

{\it Giant Eruptions:}  Another defining type of variability in LBVs is their so-called giant eruptions, which are relatively rare (these may be roughly comparable to the SN rate in a large galaxy, but there has been no quantitative rate estimate in the literature; the rate one might infer obviously depends on what one chooses to count as LBV giant eruptions vs. ILOTs, LRNe, or other transients).  What distinguishes these from the more common S Dor outbursts is that the bolometric luminosity is thought to increase substantially (usually violating the classical Eddington limit), and as a consequence, they are thought to be accompanied by tremendous mass loss in a short amount of time.  The prototypes for this sort of eruption are the famous Galactic stars $\eta$ Carinae (see Fig.~\ref{fig:etalc}) and P Cygni, which had giant eruptions in the 19th and 17th centuries, respectively \citep{sf:11,degroot:88}.  These eruptions brighten from their quiescent state by 3-4 mag at visual wavelengths, with bolometric magnitudes of $-$10 to $-$15 mag \citep{smith:11lbv}. Spectroscopy of light echoes from $\eta$ Car's 19th century eruption showed that it had an apparent temperature that was around 5,000~K \citep{rest:12}, significantly cooler than the typical value of $\sim$8,000 K seen in S~Dor outbursts.  Other claimed historical examples of these giant eruptions are SN~1954J and SN~1961V \citep{hds:99}.  It is unclear from available data if these stars actually survived the eruptions like $\eta$~Car and P Cyg did, however, with claims on both sides of the debate \citep{smith:01,smith:11lbv,vandyk:05,vdm:12,kochanek:12}.  In any case, although it was one of the two original Type V supernovae (a class introduced by \citealt{zwicky:64} that included $\eta$ Car), SN~1961V is a much more extreme object in terms of its energetics \citep{smith:11lbv}, more like SNe~IIn than LBVs, and its 1961 event is no longer considered to be an LBV giant eruption.\footnote{Its progenitor may nevertheless have been a very luminous LBV, of course \citep{goodrich:89,filippenko:95}.}  Instead, evidence and models suggest that it is more likely to be an example of a pulsational pair instability SN \citep{woosley:22}.  Some additional examples are SN~1997bs \citep{vandyk:00}, SN~2002kg \citep{weis:05,vandyk:06,maund:06}, UGC~2773-OT \citep{smith:10ip,smith:16ot}, AT~2019krl \citep{andrews:21}, and others.  An increasing number of additional candidates are being found in modern transient surveys, as noted above, although there is often some ambiguity about whether these are LBV giant eruptions or something else.  In many cases, eruptive transients  appear to have less luminous and red progenitors, as in the cases of SN~2008S and the 2008 transient in NGC~300 \citep{prieto:08s,prieto:08,thompson:09}.   

Although a star can be classified as an LBV by exhibiting either S Doradus outbursts or a giant eruption {\it a la} $\eta$ Car, there is little overlap between the two.  Despite centuries of continued study, neither $\eta$ Car nor P Cygni has shown clear S~Dor variations.   Similarly, no classical S Dor variable has shown a giant eruption like $\eta$ Car.  Are these really the same type of star tracing a similar evolutionary phase?  This remains unclear.  One possible connection between the two categories of LBV outburst is that many of the classical S Dor variables are surrounded by a massive shell nebula, which may be evidence of a past giant eruption.  Nebulae are discussed more below.

{\it New types of LBV eruptions:}  As noted above, some of the transients initially classified as LBVs in modern surveys may be other types of transients from lower-mass stars.  On the other hand, continued study of extragalactic irregular variables that are indeed luminous and blue have started to reveal additional subsets of variables or transients that don't fit neatly into the categories of S Dor outbursts or giant eruptions {\it a la} $\eta$ Car, but nevertheless appear to be related to LBVs.  Three categories of these newly recognized versions of LBVs are outlined here:

{\bf 1.  Hot LBV eruptions: Eruptions with unchanging hot temperatures.}    As noted above, both S Dor eruptions and classic giant LBV eruptions exhibit lower temperatures when in outburst, with more luminous stars tending to show very large variations in color and temperature; the main distinguishing characteristic being that S~Dor eruptions are supposed to have roughly constant $L_{\rm Bol}$ (or as noted above, relatively mild changes in $L_{\rm Bol}$), whereas giant eruptions have larger increases in $L_{\rm Bol}$.  In the past decade it has become increasingly clear that some LBV eruptions have a different type of behavior, maintaining relatively constant hot temperatures (or nearly constant) as they brighten by more than a magnitude.  This requires that they are also increasing their $L_{\rm Bol}$.  Their progenitor stars appear to be either WNH stars or Ofpe/WN9 stars, like some of the very luminous LBVs, although they maintain similar Ofpe/WN9-like spectra as they brighten in eruption, which is unlike most classical LBVs.  This doesn't quite fit into either category of S Dor outbursts or giant eruptions, although it is closer to giant eruptions.  \citet{smith:20} proposed that these might be a separate class of hot eruptions, with well-studied members including HD~5980 in the SMC \citep{barba:95,moffat:98,gloria:04,km:08,gloria:14}, GR~290 (a.k.a. Romano's star) in M33 \citep{polcaro:03,viotti:07,clark:12,polcaro:16,maryeva:19}, MCA-1B \citep{smith:20} in M33, and possibly V1 in NGC 2366 \citep{drissen:97,drissen:01,petit:06}.  One might conjecture that the hotter eruptive temperatures may result from a higher He/H abundance ratio, as the star evolves closer to the WR phase \citep{smith:20}, but this has not yet been examined with detailed models.  In any case, it has been suggested \citep{smith:20} that these hot LBV eruptions may have a connection to the relatively H-poor LBV-like eruptions that seem to precede SNe~Ibn \citep{pastorello:08,pastorello:15,smith:12hw}.

{\bf 2.  Repeating periodic and quasi-periodic LBV eruptions.} While LBVs are known for their irregular (i.e. non-periodic) eruptive variability, some do indeed show clear periodicity.  A famous example is $\eta$ Carinae, which shows dramatic events with a regular cycle of 5.52 years \citep{damineli:96,damineli:00}.  The repeating cycle manifests as a strong increase in X-ray emission \citep{corcoran:97,ishibashi:99,corcoran:00,corcoran:05,hamaguchi:14}, significant ionization changes and extended structure variability \citep{zanella:84,damineli:97,damineli:00,smith:00,smith:04purple,hillier:06,damineli:08,mehner:10,richardson:10,teodoro:12,clementel:15,weigelt:16,richardson:16,gull:21},  mild increases in optical and IR brightness \citep{whitelock:94,sterken:99,sg:00}, dust formation events \citep{smith:10dust}, and bursts of radio continuum emission \citep{duncan:95,dw:03}.  Models indicate that these events are caused by shock excitation and photoionization changes that ramp-up at periastron passes in an eccentric ($e \simeq 0.9$) colliding-wind binary system \citep{pittard:98,corcoran:01,pittard:02,okazaki:08,parkin:09,parkin:11,russell:16}.  Some of the brief bright peaks in the historical light curve of $\eta$ Car (see Fig.~\ref{fig:etalc}) also seem to coincide with times of periastron passes \citep{damineli:96,sf:11}.  These bright peaks may arise from grazing collisions or other violent interaction between the hot companion and a bloated erupting/merging star system \citep{ks:10,sf:11,smith:11eta,smith18}.  Some extragalactic SN impostors have also been seen to repeat, periodically or quasiperiodically.  SN~2000ch in NGC~3432 was named for its first outburst, which had a peak luminosity of almost $-$13 mag \citep{wagner:04}.  Even in quiescence it had a high luminosity indicative of a $\sim$60 $M_{\odot}$ star, consistent with a massive and hot LBV \citep{pastorello:10,smith:11lbv,mojgan:23ch}.  It was then seen to have a few more SN impostor outburst in 2008 and 2010 \citep{pastorello:10}. In a recent analysis of  over two decades of photometry and with more than a dozen additional eruptions,  \citet{mojgan:23ch} found that the bright recurring eruptions of SN~2000ch were periodic, with an apparent period of roughly 200 d.  \citet{mojgan:23ch} proposed that these occur because of periastron interactions in an eccentric binary, similar to the bright periastron events that occurred just before $\eta$ Car's 19th century Great Eruption.  Another example of a repeating SN impostor is AT~2016blu in NGC~4559.  Again, \citet{mojgan:23blu} analyzed a long time baseline of photometry and found many repeating outbursts that seem to favor a quasi-period of around 113 d.  In this object, the brief flashes of brightening don't occur with strict periodicity, but seem to cluster around preferred times that are periodic. \citet{mojgan:23blu} suggested that these may be similar binary interactions as in the case of SN~2000ch, but in a system with a less eccentric orbit.  

{\bf 3.  Violent pre-SN eruptions.}  The application of the term LBV to SN~IIn progenitors was mentioned above, and many examples in the literature have been reviewed by \citet{smith:14}.  Sometimes a luminous progenitor is detected directly or inferred to have a mass that is too high to be a RSG; again examples were reviewed by \citet{smith:14}.  In many cases, LBV-like precursor eruptions are inferred to have happened based on the huge mass of CSM ejected by the progenitor, required to produce the very strong CSM interaction signatures in SNe~IIn or Ibn, and especially super-luminous SNe~IIn (SLSNe~IIn).  Several previous cases are reviewed by \citet{smith:17}, with more recently analyzed examples including SLSNe~IIn like ASSASN~14il \citep{dukiya:24}, ASASSN~15ua \citep{dickinson:24}, SN~2015da \citep{tartaglia:20,smith:24} and SN~2017hcc \citep{sa:20,chandra:22,moran:23,mauerhan:24}.  Without a direct detection, it is difficult to determine if these really are LBV progenitors, or some other star that is masquerading as an LBV in every measurable way (mass-loss rate, outflow speed, H-rich composition, large mass budget, etc.).  Nevertheless, LBVs are the only observed precedent for the extreme mass loss that is required, and the inferred properties of the progenitors don't resemble any other type of known star.\footnote{One caveat is that for some of the moderate luminosity SNe~IIn, the required mass-loss rates might be achieved by some of the most extreme known RSGs like VY CMa \citep{smith:09,smith:17}.}  There are a few documented cases, however, in which the bright progenitor eruptions are detected before the SN event.  See \citet{smith:14} for previously reviewed examples.  There have also been dedicated searches for this sort of precursor behavior \citep{ofek:14,bilinski:15,reguitti:24}. Arguably the most spectacular case, and certainly the clearest connection between LBVs and SNe~IIn, is SN~2009ip.  This SN impostor was discovered in its 2009 outburst, but had archival data before that.  Unlike any other SN progenitor to date, it also had a detection of a very luminous LBV-like progenitor in quiescence a decade before its 2009 eruption, and then a 5-yr long S Dor-like outburst just before its 2009 eruption that reached a peak absolute magnitude of $-$15 mag \citep{smith:10ip}.  It then had a series of bright repeating eruptions over the next few years, which had noted similarities to LBVs like SN~2000ch and especially the brief flashes in the light curve of $\eta$ Car leading up to its 19th century eruption \citep{smith:10ip,smith:11lbv,pastorello:13,mojgan:23ch,mojgan:23blu}.  Then in 2012 it exploded as a very bright SN~IIn event, indicating that an LBV had been caught in the act of exploding as a SN \citep{mauerhan:13,mauerhan:14,smith:14}.  Although the 2012 event was also proposed to be a non-terminal eruptive transient \citep{fraser:13,pastorello:13,margutti:14}, it has since faded to be fainter than the progenitor \citep{smith:22}, and is therefore likely to have been a terminal SN.

\subsubsection{Mass loss}

The mass loss of LBVs is probably their most important property, due to the influence that their strong mass loss can have on evolution.  The steady stellar winds of LBVs in their quiescent phases are fairly well understood as standard line-driven winds of blue supergiants \citep{ln:02}.  The most extreme case is $\eta$ Car, with a mass-loss rate around 10$^{-3}$ $M_{\odot}$ yr$^{-1}$ \citep{hillier:06}.  Most other LBVs have quiescent mass-loss rates in the range 10$^{-6}$ to 10$^{-4}$ $M_{\odot}$ yr$^{-1}$ \citep{smith:14}.  The terminal wind speeds are typically in the range of 100-500 km s$^{-1}$, probably 2-3 times the surface escape velocity as for other hot supergiants.  For cooler LBVs near the bottom of the luminosity range, or for more luminous LBVs during S Dor excursions, LBVs winds are thought to be affected by the bistability jump.  This is a theoretical jump in wind driving conditions that happens at temperatures around 21,000 K due to changes in the ionization, opacity, and wind speed, leading to predicted mass-loss rates on the cool side of this jump that may increase by around a factor of 2 \citep{pauldrach:90,najarro:97,vink:99,smith:04}.  As LBVs cross this jump during an S Dor cycle, the wind speed can change (somewhat slower on the cool side), potentially leading to multiple P Cyg absorption components seen at the same time \citep{gv:11,vink:18}.  Polarization indicates that LBV winds tend to be more clumpy or aspherical than the winds of WR and O stars \citep{schulte:93,schulte:94,davies:05,wisniewski:06}.

An early idea to explain the temperature shift in S~Dor eruptions was that an increase in mass-loss rate gave rise to a pseudo-photosphere in the wind while the star's $L_{\rm Bol}$ remained constant,  with a resulting temperature around 8000 K due to opacity changes at that temperature \citep{hd:94}.   This was very reminiscent of the pseudo photosphere phase in classical novae \citep{gs:78}.  However, the pseudo photosphere explanation for S Dor events is not supported by quantitative analyses of LBV observations.  As noted above, observations indicate that $L_{\rm Bol}$ does not actually remain constant.  More importantly, quantitative spectroscopy showed that the measured mass-loss rates in outbursts do not increase nearly enough to cause a cool-enough pseudo photosphere \citep{leitherer:89,dekoter:93,dekoter:96,groh09}, and that the increasing photospheric radius is therefore more akin to a pulsation or inflation. A possible cause of this inflation of the star’s outer layers may be near-Eddington luminosities in subsurface Fe or He opacity bumps \citep{grafener:12,gl:12,jiang:15,jiang:18}. It is important to recognize that S Dor outbursts of LBVs are therefore not major mass-loss ejection events, since the S~Dor outburst mass-loss rate is only mildly elevated above that of the quiescent wind.  A possible exception to this might be R71's most recent outburst, where the mass-loss rate did increase substantially \citep{mehner:13}.  However, in this case, the $L_{\rm Bol}$ also increased substantially and the temperature in outburst was significantly lower than normal S~Dor events, so this outburst may be a very extreme case, or may actually be more like a giant eruption than an S~Dor outburst.

Giant eruptions, on the other hand, can be extreme mass-loss events.  The primary source of information about the mass ejected in LBV giant eruptions comes from the masses of shell nebulae around LBVs, while information about the outflow speed and timescales comes from the expansion speeds of nebulae, from observations of extragalactic SN impostors, and in one case, from light echo spectra and historical data.  There is a wide range of estimated masses for the shells around LBVs, with a very rough relationship between mass and the luminosity of the star \citep{so:06}.  Shell masses around $\sim$1 $M_{\odot}$ are typical for less luminous LBVs, while 10 $M_{\odot}$ is not unusual for more luminous LBVs \citep{so:06}.  On the low end, masses of only 0.1 $M_{\odot}$ are seen in the shell around P Cyg \citep{sh:06} and in the `Little Homunculus' from $\eta$~Car's 1890 eruption \citep{smith:05,ishibashi:03}.  On the high end, $\eta$~Car's bipolar Homunculus has a mass of around 20 M$_{\odot}$ \citep{smith:03,sf:07}, with some estimates as high as $\ge$45 $M_{\odot}$ \citep{morris:17}.  This mass is ejected in events lasting around a decade or sometimes much less, translating to truly extreme instantaneous mass-loss rates exceeding 1 $M_{\odot}$ yr$^{-1}$.    It is doubtful that such strong mass loss can be driven in a steady wind, suggesting hydrodynamic events instead.  If LBV mass loss is more explosive, then a mass-loss rate may not be the best way to think about them; instead, characterizing the total mass ejected and the kinetic energy may be more useful.

For LBV giant eruptions, bulk outflow velocities of 100-600 km s$^{-1}$ are typical, inferred from spectra of SN impostors \citep{vandyk:05,smith:11lbv} and from the expansion speed of young LBV nebulae like the ejecta around $\eta$ Car \citep{meaburn:87,ah:93,smith:02,smith:06,kiminki:16,smith:17pm}.  Older LBV shells often have lower outflow speeds \citep{nota:95,smith:97n,smith:97,smith:98lmc,pasquali:99,pasquali:02,danforth:01}, perhaps having been decelerated after interacting with ambient material.
Relatively fast expansion combined with massive shells points to large kinetic energy in LBV giant eruptions, in some cases with the kinetic energy significantly larger than the radiated energy in an event.  This may suggest that LBV giant eruptions have an explosive origin rather than a wind origin (see below).  In fact, some extremely high expansion speeds and direct evidence for shocks are seen.  Light echoes of $\eta$ Car's historical eruption reveal expansion speeds as high as 10,000 $-$ 20,000 km s$^{-1}$ \citep{smith:18fast}, while the distant nebulosity seen today has evidence of a blast wave seen in fast ejecta and X-rays \citep{smith:08blast,corcoran:22}.  However, this very fast material is thought to occupy a small fraction of the total mass, with most of the outflow mass at much slower speeds.  Some extragalactic LBVs also show evidence for shock excitation in their spectra \citep{smith:16ot}.

\subsubsection{Binaries, triples, etc.}

Massive stars have such a high fraction of binary systems that binary interaction likely dominates their evolution \citep{gies:86,kobulnicky:07,kobulnicky:14,sana:12,kiminki:12,demink:14,schneider:16}.  At high masses above 30 $M_{\odot}$, appropriate for LBVs, hierarchical triple and quadruple systems also become quite common \citep{moe:17,offner:23}.  Since LBVs have expanded away from the main sequence and have relatively large radii, LBVs should not have very short period companions.   Radial velocity and interferometric observations do indicate a deficit of systems with shorter periods, with longer-period binaries being fairly common \citep{mahy:22}.   

Several LBVs are indeed known to have companions in wide orbits.  As noted above, $\eta$ Car is an eccentric binary with a 5.52 yr period, identified primarily because of the observable signatures of its strong wind-wind collision, and its high eccentricity orbit may be the result of a past merger in a triple \citep{smith18,hirai:21}.  HD~5980 is in a multiple system with a relatively close WR companion and a distant O-star companion \citep{barba:95,moffat:98,gloria:04,km:08,gloria:14}.  HR Car, HD~168625, and MWC~314 each have a wide companion star that has been resolved with interferometry \citep{boffin:16,martayan:16}.  Several additional LBVs and LBV candidates, including R81, R110, MWC314, HD~326823, S~Dor, Sk -69 142a, and Sk -69 279),  have likely close companions from radial velocity or light curve studies \citep{stahl:87,vg:98,richardson:11,lobel:13,mojgan:22,mahy:22}. But because LBVs are relatively bright and their winds are variable, in many cases it is quite challenging to detect intermediate-separation companions with direct imaging or spectroscopic monitoring.  \citet{mahy:22} suggest that the lack of short-period binaries and the prevalence of wide binaries among LBVs favors a binary scenario involving mergers in binaries and triple systems rather than kicked mass gainers, since a kicked mass-gainer scenario would tend to disfavor surviving wide companions.  We direct the interested reader to a chapter about stellar mergers and common envelopes in this edition.

\subsubsection{Environments}

With the high binary fraction among massive stars, it is also important to ask what role binary interaction might play in producing LBVs, and in turn, what the consequences are for understanding single or binary star evolution.  Binary mergers and mass gainers would tend to produce very luminous, N-enriched, rotating stars that might closely resemble LBVs \citep{justham:14,schneider:24}.  A chief difficulty is that if a given star is the product of past binary interaction, then it may be quite difficult to distinguish its current properties from that of an initially very massive single star. The most straightforward way to test this is by widening one's view, and examining the surrounding stellar environment to determine if the age of the surrounding stellar population is commensurate with the presumed age of the LBV, given its luminosity and expectations from single-star evolutionary models.  This is analogous to the question of blue stragglers in lower-mass stars.

\cite{st:15} proposed a way to test this.  Massive O-type stars tend to be born in highly clustered environments, and for the most massive ones, the main-sequence lifetimes are short.  Therefore, their post-main-sequence descendants (i.e. the LBVs in a single-star view) can't have moved far from their birthsites, and they should therefore also be concentrated in young clusters.  Examining the population of LBVs in the LMC, however, \cite{st:15} found that they are not.  More than half the LBVs are in no cluster or association at all, with many of them residing more than 100 pc away from any other O-type star.  Those that were in OB associations tended to be at their periphery, and were usually associated with seemingly older populations (i.e. later O-types) than expected for their luminosity. By examining the statistical distributions of separations from O stars for 19 Magellanic Cloud LBVs, \citet{st:15} showed that LBVs are isolated from clusters of O-type stars in general, and in fact are farther away from O stars than the population of WR stars that they supposedly evolve to become.  This doesn't make sense in a single-star framework, and so \citet{st:15} proposed that LBVs are massive blue stragglers --- that they are primarily the results of mergers and mass gainers in binaries, some of which may have received a kick from a companion's SN explosion.  This would explain the apparent discrepancy between their high luminosity and their apparent isolation.  \citet{mojgan:17} showed that a passive cluster dispersal model can correctly reproduce the observed statistical spatial distribution of O-type stars on the sky, including the relative separations of early, mid, and late O-types, but cannot explain LBVs if they are the descendants of those O-type stars as expected in single-star evolutionary models.   

This new view of LBVs, that they are essentially massive blue stragglers, not only contradicts the traditional view of LBVs altogether \citep{hd:94}, but it also requires a significant reassessment of single-star models that rely upon the tremendous mass loss of LBVs to make WR stars.  The evolutionary considerations are discussed more below.  But additionally, this new result gave rise to a heated debate with some back and forth. \citet{humphreys:16} argued that LBV candidates and lower luminosity LBVs shouldn't be included in the analysis.  They excluded 16 of the original sample of 19 LBVs in the Magellanic Clouds, and argued that if one only considers the three most luminous ones (R127, R143, and S~Dor) then their statistical distribution is consistent with O-type stars after all.  However, \citep{smith:16} pointed out that one cannot get a statistically significant result with only 3 LBVs, and moreover, the median separation of those three very luminous and presumably very massive LBVs was consistent with late O-types (O8, O9) which have initial masses around 20 $M_{\odot}$.  This is still inconsistent with a single-star scenario, because those three LBVs have much higher initial masses if single.   \citet{aadland:18} also contested the results of \citet{st:15}, finding that LBVs were consistent with the separation distributions of photometrically selected bright blue stars in the Magellanic Clouds and M31/M33.  However, \citet{smith:19lbv} noted two problems with this study.  First, it was demonstrated that any photometric selection of bright blue stars is contaminated by many old blue supergiants and does not trace the distribution of O stars, and also that the angular resolution used for the M31/M33 sample was inadequate to measure small separations of O stars, causing LBVs, WR stars, and bright blue stars to all have the same separation distribution.  If a study can't measure the small separation of O stars in clusters, than it can't test whether LBVs are in those clusters either.  Thus, these two additional studies do not appear to have raised valid questions about the isolation of LBVs or the hypothesis that they are massive blue stragglers.

Two more recent studies have shed additional light on the question of LBV environments by examining their kinematics.  As noted by \citet{st:15}, two possible explanations for the isolation of LBVs are that they are merger products (making them more luminous than their brightest siblings) or that they have received a kick from a companion's SN explosion, in which case the surrounding stellar population might appear older because it is unrelated to the LBV.  In the latter case, the kinematics of LBVs should be peculiar.  \citet{mojgan:22} compared the radial velocities of LBVs in the LMC to radial velocities of RSGs, finding that LBVs have a significantly larger velocity dispersion than RSGs (40 km s$^{-1}$ as compared to 16.5 km s$^{-1}$) when one compares them to the velocities expected from the rotation curve of the LMC.  This suggests that some subset of LBVs may have indeed received a kick.  Similarly, \citet{do:24} examined the proper motions of LBVs and B[e] supergaints (which also appeared unusually isolated in the \citealt{st:15} study).  Seemingly consistent with \citet{mojgan:22}, they also found that a subset of LBVs appear to have kinematics consistent with SN kicks.  In another recent study, \citet{mehner:21} examined the stellar population close to HR Car, finding that it was associated with a moving group of B-type stars, but no early or mid O stars in this region.  HR Car is a lower-luminosity LBV with an effective initial mass (if single) around 30 $M_{\odot}$, but while no O stars around 25 $M_{\odot}$ were detected in the vicinity, the surrounding population is too sparse to determine if HR Car is a blue straggler or not using that approach.  More studies of the environments around individual LBVs may be worth pursuing.

\begin{figure}[htbp]
  \centering
  \includegraphics[width=0.85\textwidth]{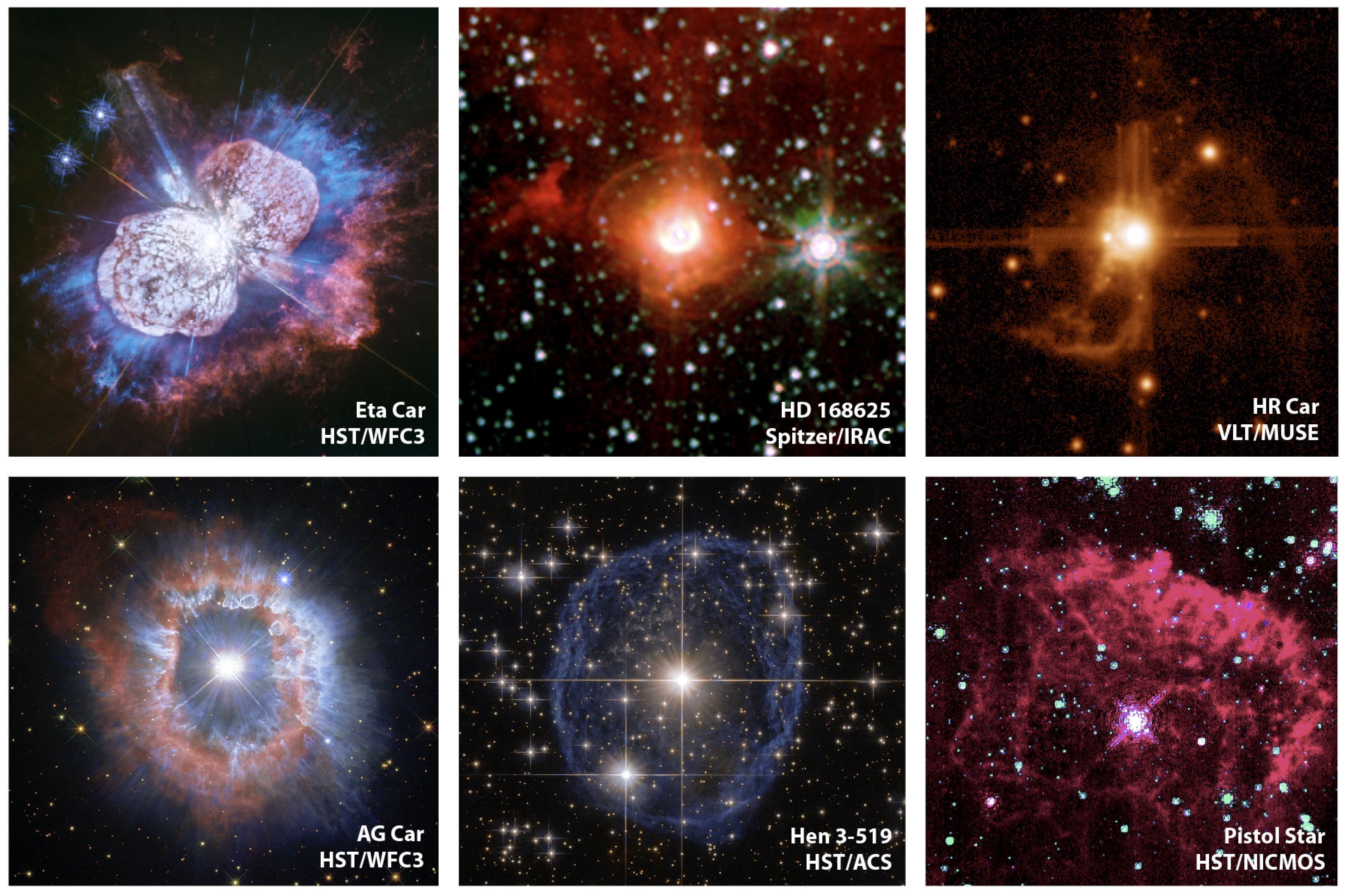}
  \caption{ 
  Gallery showing some images of nebulae around LBVs and LBV candidates, with clearly bipolar nebulae in the top row, and the more common ellipsoidal shell nebulae in the bottom row. 
  Top row: 
  {\it HST}/WFC3 image of $\eta$ Car from \citet{sm:19}. 
  {\it Spitzer}/IRAC image of HD~168625 from \citet{smith:07hd}.
  Unpublished {\it VLT}/MUSE image of HR Car made by the author from data downloaded from the ESO/VLT archive.  Interestingly, the star is not at the geometrical center of the inner nebula.
  Bottom row:
  {\it HST}/WFC3 image of AG Car (image credit: ESA/Hubble/NASA, C.\ Britt).
  {\it HST}/ACS image of Hen 3-519 (image credit: ESA/Hubble/NASA, J.\ Schmidt).
  {\it HST}/NICMOS image of the Pistol Star (image credit: D.\ Figer, NASA).
  }
  \label{fig:nebulae}
\end{figure}

\subsubsection{CSM / nebulae}

Many LBVs are surrounded by expanding ejecta nebulae that are evidence of strong mass loss in the recent past \citep{stahl:87,nota:95}.  Some examples are shown in Figure~\ref{fig:nebulae}.  As noted earlier, the total nebula masses are often quite large, typically 1-10 $M_{\odot}$ \citep{so:06}, although some have smaller or even larger shell masses.  These masses are most commonly derived by first inferring a dust mass from mid-IR SEDS, and then assuming a gas:dust mass ratio to get the total mass.  The masses, while large, are often lower limits or underestimates, unless far-IR data are available, since most of the mass can hide in the cooler dust that emits primarily at longer IR wavelengths.  Some LBVs also have mass estimates directly from diagnostics of the gas \citep{nota:95,sh:06,sf:07,bordiu:21}.   In most cases, the masses of the nebulae divided by their ages yield time-averaged mass-loss rates that are far too high to be supplied by normal winds.  If instead one divides the total mass by a duration for the mass ejection (inferred from a resolved shell thickness divided by an observed expansion speed), then the instantaneous mass-loss rates are quite extreme.  In either case, mass ejection in LBV giant eruptions is far more likely than formation by interacting winds for most of the LBV nebulae.  As noted above, mass-loss rates in normal S Dor phases are not high compared to the normal wind mass-loss rates, and nowhere near high enough to produce the massive shell nebulae seen around LBVs or to trigger significant dust formation around such luminous stars.

The most famous and well-studied LBV nebula is the complex bipolar `Homunculus' nebula surrounding $\eta$ Car (see Fig.~\ref{fig:nebulae}; top left), where proper motions have proven that it was ejected in the 19th century Great Eruption \citep{gaviola:50,currie:96,sg:98,morse:01,smith:17}.  The polar lobes exhibit linear (ballistic) expansion and recent measurements with {\it HST} give a precise ejection date of 1847, with an uncertainty of only $\pm$1 yr \citep{smith:17pm}.  This Homunculus nebula has yielded a vast literature, with hundreds of papers about the kinematics, complex multi-wavelength structure, dust and molecular chemistry, and the unusual excitation and gas-phase abundances that are too numerous to review here.  Interested readers can find a review about the Homunculus nebula itself \citep{smith:12homunc}.  There are also more recently ejected structures inside the Homunculus, including the nested `Little Homunculus' \citep{ishibashi:03,smith:05} and condensations very close to the star, both apparently ejected in the smaller 1890 eruption or soon afterward \citep{weigelt:95,davidson:97,dorland:04,smith:04acs}.  Outside the Homunculus, there is also a network of nebulosity including the newly discovered Mg~II resonance-scattering nebula \citep{sm:19,ms:24} and the complex and ratty `Outer Ejecta' \citep{thackeray:49,walborn:76,walborn:78,meaburn:87,walborn:88,ebbets:93,meaburn:96,sm:04,gomez:10,weis:12,kiminki:16,mehner:16}, with ejection dates ranging from decades to centuries before the Great Eruption \citep{kiminki:16}.  These outer features may arise from ejections during grazing collisions and the inspiral phase preceding a stellar merger \citep{smith18,hirai:21}.  Finally, $\eta$ Car is also surrounded by a soft X-ray shell and high-speed ejecta that resemble a supernova remnant \citep{chleb:84,seward:01,corcoran:04,corcoran:22}, probably arising from a blast wave from the Great Eruption \citep{sm:04,smith:08,corcoran:22}.  Altogether, this reveals an extremely complex history of recent mass loss.  It has been proposed that the Great Eruption and the surrounding nebulosity resulted from a merger of two stars in a triple system \citep{smith18}, and numerical simulations of certain aspects of the complex interaction support this suggestion \citep{hirai:21}.

Other well-studied examples of LBV nebulae include P Cygni \citep{johnson:92,barlow:94,skinner:98,meaburn:00,sh:06,arc:14}, AG~Car \citep{johnson:76,viotti:88,nota:92,smith:97,vn:15}, HR~Car \citep{umana:09,white:00,hulbert:99,voors:97,weis:97,clampin:95,hvd:91}, Hen~3-519 \citep{smith:94,stahl:87neb}, HD~168625 \citep{huts:94,rh:98,pasquali:02,ohara:03,smith:07hd,umana:10}, WRA~751 \citep{hvd:91wra,voors:00,vn:13}, the Pistol Star \citep{figer:98,figer:99}, and several others in the Milky Way \citep{nota:95}.  Some of these are shown in Figure~\ref{fig:nebulae}.  Wide-field IR imaging with {\it Spitzer} has revealed a population of massive shell nebulae around hot stars that may be LBVs or LBV candidates \citep{wachter:10,gv:10}, although further study of the central stars is warranted to confidently connect them to the class of LBVs.  Resolved LBV nebulae in the Magellanic Clouds include R127, R143, S119, S61, Sk -69 279, and others \citep{stahl:87,clampin:93,smith:98lmc,pasquali:99,danforth:01,weis:03,agliozzo:12,agliozzo:17}.  LBV nebulae are almost always found to be N rich, indicating that LBVs are evolved massive stars with CNO-processed ejecta \citep{davidson:82,dufour:89,nota:95,smith:97n,lamers:01,sm:04}.

In the admittedly biased opinion of the author, the youngest LBV nebulae (ages $<$1,000 yr), like those around $\eta$ Car and P Cygni, are the most interesting because they are still in free expansion, and therefore still retain the imprint of their ejection geometry, kinematics, and original chemical composition. Older nebulae (ages of 1,000-20,000 yr) may have already had their chemistry diluted and their kinematics altered by interaction with their ambient environment, or they may have been shaped significantly by their post-eruption wind, making their shells look more spherical or ellipsoidal.   While a few LBV nebulae are famously bipolar with tightly pinched waists, like $\eta$~Car, HR~Car, and HD~168625 (top row in Fig.~\ref{fig:nebulae}), a majority appear only mildly ellipsoidal or spherical (bottom row in Fig.~\ref{fig:nebulae}).  Generally, the older nebulae tend to appear more spherical, suggesting that their original ejection geometry may have been erased or modified.

A perhaps obvious but important point is that the older nebulae around LBVs with  ages of 1,000 to 20,000 yr indicate that some LBV giant eruptions are clearly not immediate pre-SN ejections.  LBV giant eruptions are often invoked to explain the CSM needed to power the interaction in SNe IIn, and the mass, speed, and composition of LBV nebulae are indeed very similar to the inferred CSM of SNe IIn.  However, the CSM required around SNe~IIn was ejected in the few years, decades, or centuries immediately before explosion.  While some LBVs may explode to make SNe~IIn, clearly some survive for a much longer delay after a giant eruption.  Any evolutionary scenario for LBVs should be able to account for both circumstances.

\section{Evolutionary state}

\subsection{Traditional Single-Star Scenario}

In this framework, LBVs are presumed to represent a brief transitional phase in the evolution of the most massive stars, between the main sequence O-type stars and the H-deficient Wolf-Rayet (WR) stars \citep[see many contributions in][]{nota97}.  As a massive O-type star leaves the main-sequence, it expands, it has lost mass, and its luminosity increases, presumably bringing it closer to an opacity-modified Eddington limit, and this makes it unstable.  Strong mass loss ensues until the H envelope is gone, leaving a WR star. A typical monotonic evolutionary scheme for a single star of initially 50-100 $M_{\odot}$ is \citep{langer:94}:

\smallskip 

\noindent early O star $\rightarrow$ Of/WNH $\rightarrow$ LBV $\rightarrow$
WN $\rightarrow$ WC $\rightarrow$ SN Ibc.

\smallskip

\noindent In this scenario, sometimes referred to as the ``Conti scenario" as noted earlier, LBVs are thought to be a very brief (few $\times$ 10$^4$ yr) transition when the remaining H envelope is removed by winds and eruptive mass loss, propelling the star to become a WR star and to later explode as a hydrogen-free SESN.  The heavy mass loss of LBVs, in the single-star view, is essential to remove the H envelope to form WR stars.  LBVs cannot be immediate SN progenitors in this view, because they are just about to begin core He burning and still have 0.5-1 Myr left to live.  This traditional single-star view is therefore incompatible with claims that LBVs appear to be SN progenitors, and is also in contradiction with their environments described above, where LBVs do not appear to be associated with similarly massive and young stars (if LBVs really are transitioning from core-H to core-He burning, then they should still be mostly in young clusters with O-type stars, but they clearly are not).  It therefore seems likely that this traditional single-star scenario is disfavored by some key observations of LBVs and needs to be modified or rejected altogether.  An important consequence is that if LBVs are removed from this single-star scenario, then the fates of very massive single stars also need to be revised, because without the LBV phase, a massive single star does not suffer enough mass loss to make a WR star.

For lower-luminosity LBVs, there is a different single-star scenario.  For initial masses of $\sim$25-35 $M_{\odot}$, it has been suggested that LBVs may form as post-RSGs \citep{hd:94,groh:13}.  The evolutionary trajectory would look like this:

\smallskip \smallskip 

\noindent mid/late O star $\rightarrow$ RSG $\rightarrow$ LBV $\rightarrow$
WN $\rightarrow$ SN Ibc.

\smallskip \smallskip 

\noindent The idea here is that strong mass loss as a RSG removes much of the H envelope, allowing the star to move to the blue, and causing a relatively low M/L ratio that brings it close to the Eddington limit, perhaps then initiating the LBV instability.  This scenario is not immediately in contradiction with the idea that LBVs are seen as SN progenitors, because this transition could, in principle, happen toward the end of He burning, and the star might conceivably explode as an LBV before completely shedding its H envelope to become a WR star.  However, a problem with this scenario is that it is very fine-tuned.  A post-RSG will transition to the blue only when it has almost no H envelope left \citep{meynet:15}, and so such an LBV phase would be a fleeting moment on the way to becoming a He star.  If it happened, then one would likely only see LBVs in a very narrow initial mass and luminosity range at a given metallicity, whereas the low-luminosity LBVs cover a fairly wide range \citep{hd:94,smith:04,smith:19}.  Moreover, the SNe for which LBVs are invoked as progenitors are SNe IIn, which are required to shed several $M_{\odot}$ in their final moments --- but an LBV that is a post-RSG does not have that much H left.  Instead, such blue-transitioning stars would likely appear as SNe~IIb  \citep{heger:03}.  A bigger problem with this scenario is that the observed winds of RSGs are simply nowhere near strong enough to remove the H envelope for initial masses below about 25 $M_{\odot}$ \citep{beasor:20,beasor:21}.  Perhaps this sort of scenario might still work for initial masses of 25-35 $M_{\odot}$, if their RSG wind mass loss is strong enough; this remains unclear.   Instead of normal winds, perhaps a better way  to make an LBV as a post-RSG is via binary RSGs that enter a common envelope phase or experience RLOF mass stripping.  Indeed, studies of SNe suggest that the vast majority of SESNe arise from binary mass transfer at relatively low initial masses \citep{smith:11frac,drout:11,sana:12}, not single-star mass loss.

\subsection{Binary Evolution Scenario}

The fact that LBVs are more isolated than WR stars, plus the fact that some stars appear to remain in an LBV phase until death as SNe~IIn, suggests an entirely different picture.  Instead of the monotonic evolutionary scheme for single stars, the spatial distributions of LBVs suggest that massive star populations and especially LBVs are dominated by bifurcated evolutionary trajectories \citep{st:15}:

\smallskip
\noindent O star binary $\rightarrow$ $\begin{cases} {\rm WN} \rightarrow
  {\rm WC} \rightarrow {\rm SNIbc} & {\rm (donor)} \\ {\rm LBV/B[e]}
  \rightarrow {\rm SNIIn} & {\rm (gainer \, or \, merger)} \end{cases}$.

\smallskip

\noindent In this scenario, O-type stars evolve off the main sequence, and through binary interaction the majority of massive stars either (1) lose their H envelope through mass transfer to a companion, become WR stars, and then die as stripped-envelope SNe, or (2) gain mass from a companion through RLOF or merger, become BSGs, sgB[e]s, and LBVs, and then retain their H envelopes until they die as SNe~IIn or other varieties of SNe~II. These mass gainers would be overluminous compared to surrounding stars, they would get spun up, and probably enriched in products of CNO burning.  These are all observed traits of LBVs.  Similar to mass gainers, many of the same observed results would occur if LBVs are the products of binary mergers \citep{justham:14}.  Kinematic evidence discussed above suggests that LBVs may be a mix of both kicked mass gainers and merger products \citep{mojgan:22,do:24}.  This sort of mass-gainer or merger scenario for LBVs \citep{st:15} was mentioned early on by \citet{kg:85}, but seemed to be disregarded or largely ignored until recently.  Evidence from the environments of LBVs, discussed above in section 2.25, seems to rule out a simple single-star scenario and strongly favors a scenario where LBVs are largely the products of past binary interaction as mass gainers or mergers.  Yet, specific scenarios involving past interaction can be tricky to prove, and additional work is needed on LBV environments, plus theoretical work on the complex evolution of mass gainers and merger products.

\section{Instability and Eruptions}

Ever since \citet{herschel:1847} described the sudden flashes and relapses of $\eta$ Argus (a.k.a. $\eta$ Carinae), the irregular eruptive variability of LBVs has been both captivating and difficult to understand. Early ideas centered on a geyser-like model for LBV  eruptions \citep{maeder:92}, where very luminous stars reach cool temperatures in their post-main-sequence evolution, allowing a recombination front (akin to a boiling front in a geyser) to proceed into the star, thereby initiating a rise in mass-loss because of the change in opacity.  Maeder's suggestion was adopted by \citet{hd:94}, who referred to LBVs as ``astrophysical geysers".  However, this scenario is disfavored or even ruled out by most observational evidence.  For the giant eruptions, there is simply not enough energy or mass in the outer envelope of the star that would experience this sort of instability, so this scenario fails to account for the large observed mass and kinetic energy budgets of these events. A hypothetical geyser scenario doesn't work for the more traditional S~Dor outbursts either, because as noted above, observations show that they are not major mass-loss events with extreme pseudo photospheres; instead they are more akin to temporary envelope inflation or pulsation \citep{leitherer:89,dekoter:93,dekoter:96,groh09}.  That sort of inflation of the outermost layers may be achieved by opacity changes in sub-surface layers of a star that is not too far from the classical Eddington limit.  This idea that an opacity-modified Eddington limit or runaway pulsations may be responsible for the normal S Dor variations of LBVs has been around for a long time \citep{lf:88,gk:93,sc:96,lamers:97,guzik:99,stothers:99}, and recent work has made significant advances on this front \citep{grafener:12,owocki:13,owocki:15,grafener:15,jiang:15,jiang:18,grass:21,agrawal:22,cheng:24}.  The lower effective gravity related to rapid rotation, working in tandem with opacity in the cooler outer layers, may have a similar influence.  This is sometimes referred to as the $\Omega$ limit (or the $\Omega$-$\Gamma$ limit), to emphasize the contribution of rotation to the instability in an LBV's outer layers \citep{langer:97,langer:98,glatzel:98,langer:99,mm:00}.  One potential difficulty for this type of opacity-modified Eddington limit scenario is that a number of the less luminous LBVs have quite modest luminosities that do not approach the Eddington limit \citep{smith:19}, so these LBVs would require a different mechanism, even though they exhibit similar irregular instability.

The giant eruptions have been more difficult to understand.  Due to their observed large increases in $L_{\rm Bol}$, the most common theoretical mechanism investigated for LBV giant eruptions has been super-Eddington winds.   Although the idea is fairly straightforward (the force of radiation overwhelms the force of gravity, leading to runaway mass loss driven by the continuum opacity of electron scattering), interesting complexities arise including the development of porous atmospheres that allow a prolonged steady state \citep{owocki:97,shaviv:98,shaviv:00,shaviv:01,owocki:04}, as well as high optical depths that enter the regime of photon tiring and winds that may partly stall or even fall back to the star \citep{owocki:04,owocki:08,vanmarle:08,vanmarle:09,quataert:16,owocki:17}.  While super-Eddington winds as a mass-loss mechanism are well motivated by the observed fact that LBV giant eruptions do exhibit large increases in luminosity, invoking a super-Eddington wind does not provide a reason why the star becomes super-Eddington in the first place, and there is no proposed source for the extra energy.  So far, there has been no proposed mechanism for a single star in transition from H to He core  burning to quite suddenly increase its $L_{\rm Bol}$ by a factor of 5 to 10 or more.

Several lines of evidence have suggested that giant eruptions may be more like hydrodynamic explosions than winds, and the evidence in favor of an explosive mechanism is quite clear in the case of $\eta$ Car's Great Eruption.  Considering the total mass and expansion speed of the Homunculus nebula, the kinetic energy is around 10$^{50}$ erg,  whereas the total radiated energy in the Great Eruption is only about 2$\times$10$^{49}$ erg.  Winds tend to have very low ratios of kinetic energy to radiated energy, whereas SNe tend to be mostly kinetic energy.  $\eta$ Car is somewhere in between, but leans more toward explosions.  In addition, the kinetic energy budget more than doubles if we include very fast ejecta outside the Homunculus \citep{sm:04,smith:08blast} and the kinetic energy that was required to make the X-ray shell \citep{corcoran:22}.  The extremely thin walls of the Homunculus require shock compression and radiative cooling \citep{smith:13}. And finally, light echoes of $\eta$ Car revealed that a small fraction of the outflow was accelerated to extremely high speeds of 10,000-20,000 km s$^{-1}$ during the Great Eruption; this requires shock acceleration \citep{smith:18fast}.  Some extragalactic LBV-like transients have also shown very high outflow speeds and evidence of shock excitation \citep{smith:10ip,pastorello:13,smith:16ot}.

One potential mechanism for a non-terminal explosion in a massive star is the pulsational pair instability (PPI; see the chapter on the pair instability in this volume for more information; Renzo \& Smith, this volume).  \citet{woosley:17} pointed out that PPI models might do a good job of matching the erratic long-duration light curve, the mass and energy budgets, and the fast ejecta seen in $\eta$ Car.  However, this model also predicts several other observables that seem inconsistent with LBVs \citep{woosley:22}.  Moreover, the PPI is restricted to very massive stars with initial masses above $\sim$95 $M_{\odot}$, so while it might work in a very massive star like $\eta$ Car, it could not explain the giant eruptions seen in most of the other LBVs that have initial masses down to around 20 $M_{\odot}$ \citep{smith:04}.  Since the pair instability usually kicks-in during O burning shortly before a star's demise, it would also have a hard time accounting for many LBVs that have had a giant eruption 10,000-20,000 yr ago (based on the age of their nebulae), but have been quiescent since.  \citet{hw:02} did note a rare case where the PPI eruption can delay the resumption of nuclear burning, leading to intervals of as much as
1,000 yr between bursts, but this is not true over much of the initial mass range susceptible to the PPI.

Given the high fraction of binaries among massive stars, and the environmental evidence that points to LBVs being a binary phenomenon, it is perhaps surprising that there has been so much interest in single-star models for LBV eruptions.  Some binary scenarios have been considered, though.  In fact, binary scenarios were discussed early on \citep{kg:85,jsg:89}, and in hindsight seem quite plausible.  These seem to have been discounted throughout the 1990s by most researchers as a matter of preference, rather than because of compelling counter-arguments. Qualitative comments about the stark isolation of LBVs were also highlighted long ago \citep{lortet:89}, but were apparently overlooked.  More recently, binaries have been increasingly discussed in connection with LBVs, though. In a series of papers, Soker and collaborators have advocated a type of binary model for LBV-like eruptions and other related transients that are powered by accretion onto a companion star \citep{soker:07,hs:09,ks:10,ks:10b,ks:16,soker:16,soker:20}.  In these models, accretion onto a main-sequence star companion at close passages during periastron is invoked to account for the extra luminosity seen in light curves, while the jet outflows may explain the bipolar shapes of nebulae and the presence of fast moving material.  These models typically invoke a large burst of mass loss from the primary star in order to provide the mass that is then accreted at a high rate onto the secondary.  Of course, $\eta$ Car is known to be an eccentric binary system, and \citep{smith:11eta} pointed out that with the radius of the photosphere required to account for the observed luminosity during the Great Eruption, it was inevitable that the companion would plunge inside the bloated envelope of the erupting star (or the common envelope of two stars).  The times when periastron collisions would occur coincide with brief flashes in the light curve, as noted earlier.  But in terms of powering the Great Eruption itself, a merger of two massive stars seems like the most straightforward source of ejected mass and energy.  \citet{smith18} proposed that a merger of two stars in a triple system, while admittedly a complicated scenario, was the only model that could explain many different observables of the $\eta$ Car system, including the present day WR companion on a wide eccentric orbit.  In Smith et al.'s  model, grazing collisions before the merger caused the ragged outer ejecta, and the final inspiral to a merger created a torus that then pinched the waist of the merger ejecta, creating the bipolar nebula we see today.  Numerical models support aspects of this merger scenario \citep{hirai:21}.  Similar models have been proposed for less massive stellar merger transients \citep{pejcha:16,pejcha:17,mp:17}.  \citet{justham:14} have argued that the observed properties of LBVs are consistent with being merger products, and as discussed above, the surrounding environments that make LBVs stand out as massive blue stragglers also favor the idea that LBVs are merger products.  While post-main-sequence mergers of massive stars are undoubtedly exceptional events, it is also true that LBVs are quite rare.

\section{Conclusions and Outlook}\label{sec:conclusions}

Since the term LBV was introduced 40 years ago, this class of stars has been difficult to define and to understand. This may because they are rare and each one seems unique at some level.  As we continue to study them, we seem to find more examples that don't quite match some early narrow definitions that were based on just a few objects. For example, under some definitions of an LBV, it is hard to know what to do with the new subset of LBV giant eruptions that brighten at constant hot temperature, or the pre-SN outbursts that look like LBVs in every measurable way, but are clearly not in transition to a prolonged WR phase.  Since we still don't understand the mechanism of LBV outbursts and eruptions, it is probably wise to be inclusive of the cases that don't fit established ideas in every way, because they may help inform the broader phenomenon of eruptive instability in massive stars. This may be better than only including the objects that fit preconceived ideas; inclusivity seems to have been the original spirit of the umbrella term `LBV' introduced by \citet{conti:84}.

At present, the key questions about LBVs seem to center on their role in evolution --- whether they adhere to the traditional view of very massive single stars in transition to the WR phase, or if instead they are massive blue stragglers that are products of past binary interaction.  As noted in this review, mounting evidence seems to require the latter, although this remains difficult to prove and even more difficult to understand physically.  It may be useful to consider the question from a different angle.  Starting from the perspective that binary and triples do indeed exist and are now known to be very common among massive stars, we should expect that there will be cases of mass gainers and post-main-sequence mergers.  It is fair to ask what such binary products should look like (overluminous blue stragglers, rapid rotators, He and N-rich, surrounded by very massive shells from a merger event, probably quite rare,  etc.), and then ask which observed classes of stars would best fit that description.  LBVs seem to check all the boxes, and seem like the best candidates among known classes of massive stars at high luminosity. The B[e] supergiants seem like excellent candidates at somewhat lower luminosities \citep{ppod06,st:15,wu:20}.  LBVs seem consistent with expectations of binary products in most ways, even if it may be difficult to prove for any individual object.  As such, it seems likely that LBVs, their instability, and their mass loss may be among our best resources for obtaining clues about the physics of binary mass transfer and merger events among the most massive stars.

With {\it JWST}, Rubin, and other facilities operating, it is virtually guaranteed that the coming decade will provide many additional examples of weird extragalactic non-SN transients or SN precursors that resemble LBVs, but that don't exactly match our notions of what LBVs are supposed to look like.  Continued studies of the detailed physical parameters of LBVs and their nebulae will also be useful.  $\eta$ Car has been a gold mine for observational clues about erupting massive stars, but it is after all just one object; it would be useful to have comparable information for other objects too.  Additionally, continued detailed studies of the environments around LBVs will be useful; not only to determine if they are inconsistent with single-star evolution expectations, but to more accurately quantify any age discrepancies between the surrounding stars and the LBV itself (i.e. the degree to which each LBV is a blue straggler).  This may help us better understand how much mass is gained in RLOF, or the range of mass ratios that lead to mergers, as compared to the amount of mass ejected in merger events.  These are important inputs for understanding the physics of such events.

The area that probably holds the most potential for progress in the next decade is theoretical expectations for what mass gainers and mergers should look like, and what the unstable short-duration events themselves should look like, so that these predictions can be more meaningfully compared to the observed properties of LBVs.  There is currently a dearth of model predictions to compare to observations, both in terms of individual objects over a range of stellar parameters, and also in terms of expected populations.  This is a hard problem, as the detailed physics of mass transfer, common envelopes, and mergers is difficult anyway; more so for very massive stars.

\begin{ack}[Acknowledgments]%

NS is grateful for many enlightening discussions about LBVs, binary interaction and evolution, mass loss, eruptions, and explosions over the years, including those with D.\ Arnett, S.\ de Mink, Y.\ G\"otberg, S.\ Justham, N.\ Langer, S.\ Owocki, P.\ Podsiadlowski, E.\ Quataert, M.\ Renzo, J.\ Vink, and S.\ Woosley.  Some support was provided by the National Aeronautics and Space Administration (NASA) through HST grants GO-13787, AR-14316, and GO-16649 from the Space Telescope Science Institute, which is operated by AURA, Inc., under NASA contract NAS5-26555.

\end{ack}

%\seealso{ \cite{marchant:19} provide at
%  \url{https://zenodo.org/records/3786599} movnot covered
%  above.}

\bibliographystyle{Harvard}
\bibliography{lbv-ref}

\end{document}